# Effect of rapid thermal annealing on short period {CdO/ZnO}$_m$ SLs grown on *m*-Al$_2$O$_3$

A. Lysak, E. Przeździecka, R. Jakiela, A. Reszka, B. Witkowski,
Z. Khosravizadeh, A. Adhikari, J.M. Sajkowski, A. Kozanecki

Institute of Physics, Polish Academy of Sciences, al. Lotnikow 32/46, Warsaw, Poland

Here, we report on the characterization of {CdO/ZnO}$_m$ superlattice structures (SLs) grown by plasma assisted molecular beam epitaxy. The properties of *as-grown* and annealed SLs deposited on *m*-oriented sapphire were investigated by secondary ion mass spectrometry (SIMS) and scanning electron microscopy (SEM) in cathodoluminescence (CL) and energy dispersive X-ray modes. The deformation of the crystallographic structure of SLs was observed after rapid thermal annealing at 900°C in oxygen flow due to migration and segregation of Cd atoms. SIMS measurements revealed that the distributions of cadmium in the annealed samples depend on the thicknesses of the CdO and ZnO sublayers in the *as-grown* superlattice structures. Depth-resolved CL measurements showed that shifting of the near band edge emission peaks is closely related to the Cd profiles measured with SIMS.

## I. Introduction

Transparent conducting oxides (TCOs) form a special class of semiconductors, which have wide band gaps, low resistivity and high transparency in the spectral region from ultraviolet to visible. TCOs materials include, for example, In$_2$O$_3$, ZnO, CdO, SnO$_2$, Fe$_2$O$_3$, TiO$_2$ and many other binary and ternary oxides [1–3]. Growing interest in the research of TCO structures is due to their potential applications in optoelectronics, photovoltaics and nanoplasmonics [4–6].

Zinc oxide (ZnO) is a II–VI semiconductor with the wurtzite crystal structure [7] and a wide band gap in the near UV range ($E_g$ ~ 3.31 eV at room temperature). One of the very important advantages of ZnO is the high binding energy of free excitons (60 meV), which allows to generate efficient emission at room temperature. Due to its relatively low cost, this material is used in many optoelectronic devices [8–12]. Band gap engineering by alloying ZnO with other metals in a wide range of compositions is attractive for practical applications. Actually, ZnO-based heterostructures, such as Zn$_{1-x}$Mg$_x$O [13,14], Be$_x$Zn$_{1-x}$O [15,16], Zn$_{1-x}$Ca$_x$O [17,18] and Zn$_{1-x-y}$Be$_x$Mg$_y$O [19], which shift the band gap into the deep ultraviolet range, are intensively investigated. On the other hand, alloying ZnO with CdO reduces the band gap [20]. It opens some possibilities to deepen quantum wells in multilayer structures containing ZnCdO layers. Moreover, thin-film transistors [21] and UV detectors [22] based on MgZnO/CdZnO heterojunctions have recently been developed with success [23].

The theoretical [24,25] and experimental [26,27] data demonstrated the changes of the ZnCdO alloy energy gaps within the limits of 3,3 - 1,8 eV with increasing Cd concentration. The red shift of the optical band gap in Zn$_{1-x}$Cd$_x$O alloy is due to the fact that CdO has a direct ~2.35 eV and two indirect (1.28 and 0.8 eV) band gaps [28].

The synthesis of Zn$_{1-x}$Cd$_x$O thin films is possible by various methods, such as pulsed laser deposition (PLD) [29,30], spray pyrolysis [31–33], sol–gel [32,34], reactive magnetron sputtering [26,35,36] and molecular beam epitaxy (MBE) [35,36], to mention the most commonly used technologies. MBE technology is well suited for growing ZnO-based SLs at low temperatures as it allows to control layer thicknesses with the precision of one atomic layer. Regardless of the method of growth, there are several limitations that affect the possibility of obtaining high-quality $Zn_{1-x}Cd_xO$ heterostructures with high Cd content [38]. The most important reason is the different crystal structure of CdO and ZnO. The rocksalt structure of CdO ($\alpha = 4.70$ Å) limits its equilibrium molar solubility in the wurtzite ZnO (a = 3.25 Å, c = 5.21 Å) to 2% [39]. Gang Yao et al. [30] showed that it is possible to obtain single-phase $Zn_{1-x}Cd_xO$ layers with a Cd content of x = 9.60 at. % by the PLD method. The multiphase or polycrystalline nature of the material seriously complicates the possibility of obtaining good quality thin films, and heterojunctions, as well as QWs and superlattices (SLs) based on $Zn_{1-x}Cd_xO$ alloys. Another reason is the high volatility of Cd element, which prevents the use of high growth temperature in the synthesis of structures containing $Zn_{1-x}Cd_xO$ layers. Nevertheless, K. Sakurai et al. [40] reported the possibility to produce $Zn_{1-x}Cd_xO$ thin films with a small Cd content by MBE also at higher temperatures (400 – 600 °C).

Although the optical, electrical and crystallographic properties of $Zn_{1-x}Cd_xO$ heterostructures have been intensively studied, there exist only a few reports about their thermal stability [41–43]. The research of these alloys was focused mainly on the optical properties of thin films, however, information about the distribution of cadmium atoms after annealing is generally lacking.

In order to avoid phase separation in random ZnCdO layers, artificial alloys in the form of short-period CdO/ZnO SLs, named also quasi-ternary alloys, can be used e.g: ZnO/CdO or ZnO/ZnCdO [44]. Such structures have been studied very rarely, and currently only one publication, by I. Suemune at el. [45] is known, in which the ZnO/CdO SLs were obtained by metalorganic molecular-beam epitaxy (MOMBE).

In this work, we report on the growth and annealing of the $\{CdO/ZnO\}_m$ SLs deposited on $m$-$Al_2O_3$ substrates by the MBE method. The uniqueness of this work lies in the analysis of the depth distribution of Cd atoms in *as-grown* SLs and after annealing them at high temperature by secondary ion mass spectrometry (SIMS), and then correlating the SIMS profiles with the results of energy dispersive X-ray (EDX) emission and spatially resolved cathodoluminescence (CL) measured using a scanning electron microscope (SEM).

**Experimental details**

The set of $\{CdO/ZnO\}_m$ SLs was grown on (10-10) $m$-plane sapphire substrates by plasma assisted molecular beam epitaxy (PA-MBE) in a Riber Compact 21 system. Two series of structures A and B, each containing 3 samples, and differing in the first deposited layer from which the structures began (CdO or ZnO). The SLs in a given series also differed, for example, in the thickness of the sublayers and/or the number of CdO/ZnO pairs (Table 1).

Before the growth, the $m$-$Al_2O_3$ substrates were chemically cleaned in a (1:1) $H_2SO_4$:$H_2O_2$ mixture for 5 min and then rinsed in deionized water and dried with nitrogen gas. Thermal purification of the substrates was carried out at a temperature of 150 °C for 1 hour in the load chamber and then the substrates were transferred to the growth chamber where they were cleaned in vacuum at 700 °C for 10 min and then for 30 min in oxygen plasma also at 700 °C. During the growth process, the radio-frequency (RF) power of the oxygen plasma was fixed at 400 W, and pure 6N $O_2$ gas flow rate was 3 sccm. All $\{CdO/ZnO\}_m$ structures were grown at a temperature of 360 °C measured using a thermocouple located close to the substrate. Pure Zn and Cd 7N metals was evaporated from conventional two-zone Knudsen cells. The fluxes of Cd and Zn were measured before the growth with a beam flux monitor of the Bayard-Alpert type ionization gauge and they were

~1,5e-7 torr and 9e-7 torr for Cd and Zn, respectively. The Cd and Zn fluxes were fixed the same for all growth processes. The growth rate was previously determined [46] by measuring the layer thicknesses with a transmission electron microscope (TEM) and was about 1.5 nm/min for both layers. Based on TEM analysis (not shown here) the SLs thicknesses were in the range of 225-337 nm.

Post-growth thermal treatment was performed in a rapid thermal processing (RTP) system (AccuThermo AW610 from Allwin21 Inc.) at 900°C in an oxygen ($O_2$) atmosphere by 5 minutes.

The depth profiles of the Cd, Zn and O elements in $\{CdO/ZnO\}_m$ SLs were investigated by secondary ion mass spectrometry the SIMS method using a CAMECA IMS6F system. A primary Cs+ ion beam at an energy of 5.5keV and a current of 50 nA was used. The Cs+ ion beam covered the area of 150 μm ×150 μm and the secondary ions were collected from the central region with a diameter of 60 μm. The signals from the $^{114}$CdCs+, $^{64}$ZnCs+ and $^{16}$OCs+ clusters were analyzed.

A Hitachi SU-70 scanning electron microscope (SEM) coupled with a Thermo Fisher Scientific Energy Dispersion X-ray (EDX) spectrometer with a Li-drift silicon X-ray detector and the Noran System7 was used to study the morphology of *as-grown* and annealed $\{CdO/ZnO\}_m$ SLs. The microscope was also equipped with a Gatan MonoCL3 cathodoluminescence (CL) system and a liquid helium cooled cryo-stage with which the local optical properties of SLs (~ 5 K) were investigated at low temperatures. To perform a depth-profiling analysis, the CL was excited by an incident electron beam at an accelerating voltage (AV) varied from 2 to 15 kV, in order to vary the electron beam penetration depth. The beam current was varied from 0.164 nA to 1.23 nA. The beam current and its energy were set in such a way that the excitation power was the same for each selected energy. To estimate the depth from which the CL signal was generated for the particular AV, Monte Carlo simulations were performed using Casino v2.48 software [47].

**Results and discussions**

It is known from the previously reported data [46] that the SLs structures analyzed in this work have alternately cubic (CdO) and hexagonal (ZnO) structures. The thicknesses of individual ZnO and CdO layers were also estimated based on the TEM results [46] and these data are presented in Table 1.

The depth profiles of Cd, Zn and O in *as grown* $\{CdO/ZnO\}_m$ SLs measured by SIMS are presented in Fig. 1 (a-f). It can be seen, that the individual CdO and ZnO layers are clearly traceable and their order reflects the planned structures. The analysis of the SIMS depth profiles made it possible to estimate the [Cd]/[Zn] ratio using the signal intensities of the $^{114}$CdCs+ and $^{64}$ZnCs+ ion clusters. Since only the signal intensities from the clusters containing the most abundant isotopes of Cd and Zn were analyzed, the calculations required to take into account abundances of 0.287 and 0.486, respectively. The interpretation of the SIMS results is influenced by the so-called "matrix effect" which reduces the accuracy of the estimation of Cd and Zn contents [42]. That is why a reliable determination of the composition of a ternary Cd compound and multilayer structures using the SIMS method requires additional measurements on the appropriate standard samples [48].

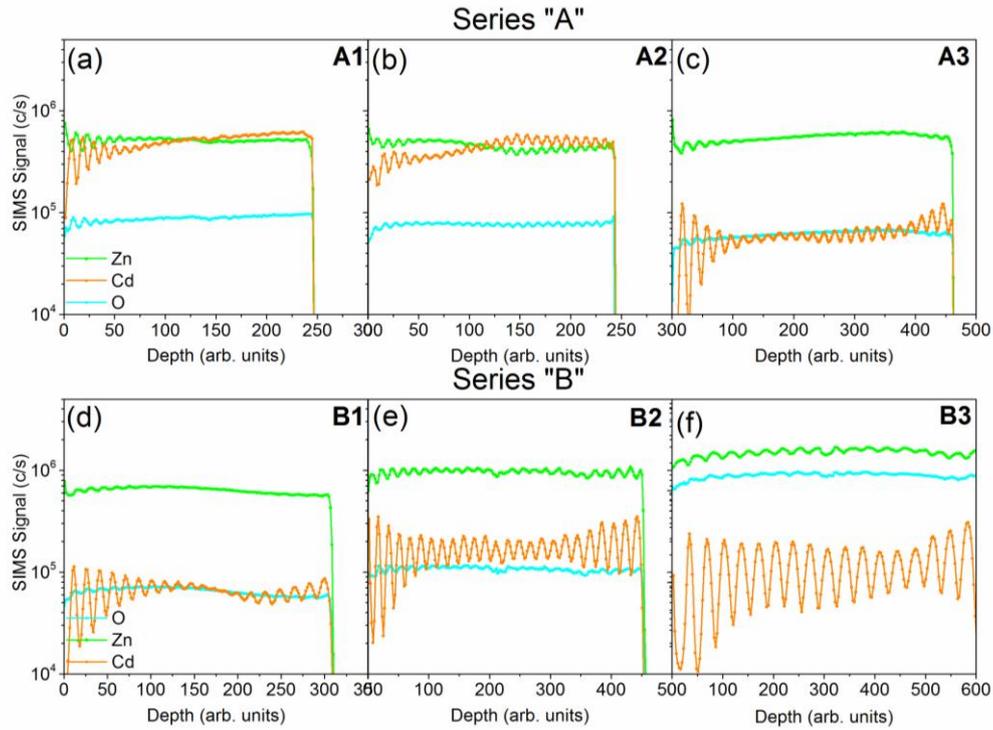

**Fig. 1.** SIMS depth profiles of Zn, Cd and O elements in *as-grown* SLs: (a) $\{CdO_{12nm}/ZnO_{1.5nm}\}_{25}$; (b) $\{CdO_{9.3nm}/ZnO_{4.5nm}\}_{18}$; (c) $\{CdO_{5nm}/ZnO_{5nm}\}_{25}$; (d) $\{ZnO_{8nm}/CdO_{1nm}\}_{25}$; (e) $\{ZnO_{9nm}/CdO_{5nm}\}_{18}$ and (f) $\{ZnO_{5nm}/CdO_{5nm}\}_{25}$.

**Table 1.** Summary data for *as-grown* $\{CdO/ZnO\}_m$ superlattices.

| Sample | Thicknesses of CdO/ZnO layers, nm | Number of layer pairs | Cd at.% (EDX) | Cd/Zn (EDX) | Cd/Zn (SIMS) | Cd at % (SIMS) |
|---|---|---|---|---|---|---|
| A1 | 12/1.5 | 25 | 38 | 0.62 | 1 | 0.5 |
| A2 | 9.3/4.5 | 18 | 46 | 0.86 | 0.95 | 0.49 |
| A3 | 5/5 | 25 | 9 | 0.10 | 0.11 | 0.1 |
| B1 | 8/1 | 25 | 8 | 0.10 | 0.1 | 0.09 |
| B2 | 9/5 | 18 | 14 | 0.16 | 0.18 | 0.15 |
| B3 | 5/5 | 25 | 7 | 0.08 | 0.08 | 0.07 |

The average chemical composition of the $\{CdO/ZnO\}_m$ SLs was examined by EDX using 6 keV electrons to collect the signal from the entire depth of the SLs and not to excite the substrates. The most intense peaks were observed for the Zn Lα, Cd Lα, and O Kα transitions. The [Cd]/[Zn] ratio as derived by the EDX method was compared (Table 1) to the CdCs+/ZnCs+ SIMS signal ratios (Fig. 2). As can be seen, the 1:1 proportionality is maintained for low, below 0.2, Cd content. For two samples with higher Cd-to-Zn ratios the results obtained by these two methods differ considerably. Such discrepancy may be assigned to the so-called "matrix effect" in SIMS technique.

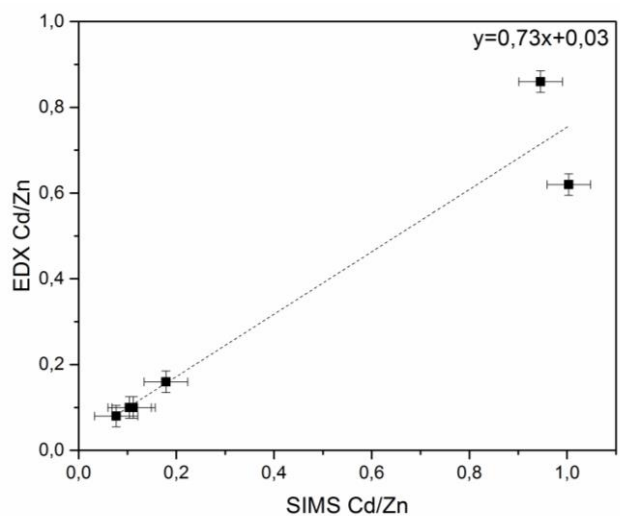

**Fig. 2.** Comparison of SIMS CdCs$^+$/ZnCs$^+$ signals ratio to EDX data of [Cd]/[Zn] ratio for *as grown* SLs.

To study the thermal stability, the structure series "A" and "B" were RTP annealed at 900 °C for 5 min in an O$_2$ atmosphere. The effect of high temperature annealing on the distribution of Cd in the samples was investigated by the EDX method under the same conditions as the *as-grown* SLs and the results are shown in Table 2. It can be noticed that the average Cd content has decreased compared to its content in *as-grown* structures. The decrease in the Cd content may be associated with the evaporation of Cd atoms through the surface. Azarov et al. [43] estimated the average evaporation rate at 900 °C annealing in air to be ~ 3·10$^{15}$ Cd/(cm$^2$·s), that means which indicates a rapid migration of Cd atoms at high temperatures.

**Table 2.** Summary data for annealing of {CdO/ZnO}$_m$ superlattices at in an O$_2$ atmosphere.

| Sample | Annealing temperature (°C) | Cd at.% (EDX) | Cd/Zn (EDX) |
|--------|----------------------------|---------------|-------------|
| A1     | 900                        | 18            | 0.23        |
| A2     | 900                        | 17            | 0.21        |
| A3     | 900                        | 4             | 0.04        |
| B1     | 900                        | 3             | 0.03        |
| B2     | 900                        | 4             | 0.05        |
| B3     | 900                        | 3             | 0.03        |

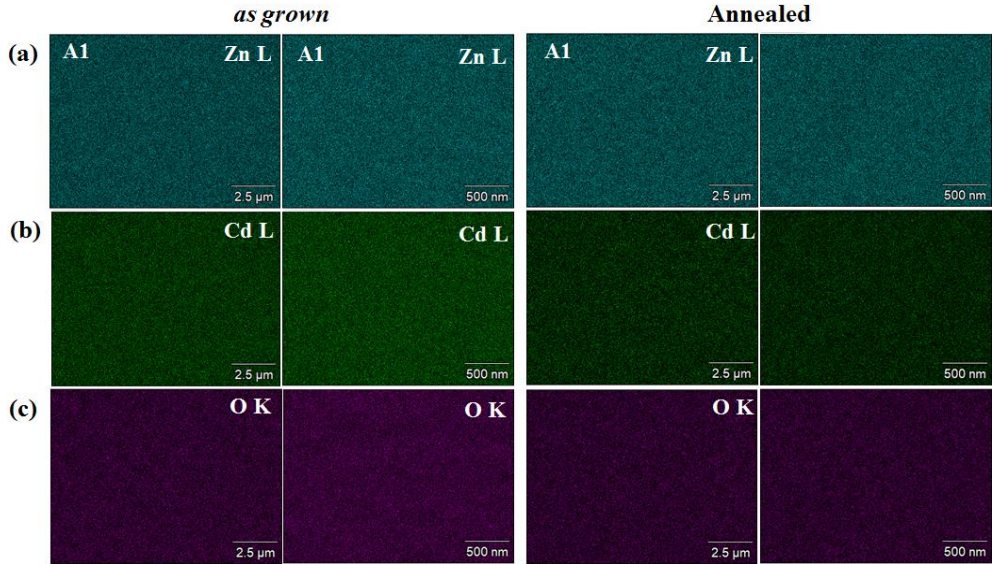

**Fig. 3.** Images showing elemental distribution from EDX; (a)–(c) show the distribution of cadmium, zinc, and oxygen elements, respectively. Samples *as grown* and annealed was presented on the left and right column, respectively.

The *as grown* and annealed layers were further investigated by EDX mapping technique resulting in elemental distribution maps. Figs. 3 a–c show the elemental distribution maps of the selected sample measured at two magnifications. The green, blue, and pink colors are assigned to the distribution of cadmium in Fig. 3b, zinc in Fig. 3c, and oxygen in Fig. 3d elements, respectively. As was shown, spatial distribution of Cd, Zn and O is homogenous. Based on the Monte-Carlo simulations [47] for 10 kV accelerating voltage the radius of the EDX signal generation area in the samples is 15 nm, so it would be possible to observe precipitations larger than 30 nm. Therefore we do not see any lateral inhomogeneity of Cd in our samples.

To study the distribution of Cd atoms after annealing in detail, two experimental techniques were used, namely SIMS and in-depth profiling of CL at low temperatures. CL in-depth profiling is an interesting method that allows to visualize the optical properties of samples at different depths. As mentioned above, this method is based on increasing the AV (energy of the electron beam) used to excite the CL. As a result of increasing the energy, layers located deeper and deeper contribute to the CL spectra. Therefore, it is necessary to perform Monte-Carlo simulations to take into account the spatial distribution of electrons in the material.

Monte-Carlo simulations of the depth dependent CL generation for various AVs and the electron beam exciting the sample from the top were done with the Casino v 2.4.8.1 software [47]. A thick $Zn_{1-x}Cd_xO$ layer with x=10% cadmium content was used for the simulations as an approximation of the studied structures. The simulations were performed for 1000 electrons and the beam radius of 3 nm. The normalized CL generation dependence on the applied AV is shown in Fig. 4. Various AVs (AV = 2, 6, 10 and 15 kV) were proposed to study sample homogeneity as a function of depth in the SLs. The CL peak intensities for energies of 2, 6, 10 and 15 keV correspond to depths of ~ 15, 75, 180 and 330 nm, respectively. However, it must be remembered that the entire CL signal comes from the total excited volume of the sample, which increases with increasing AV.

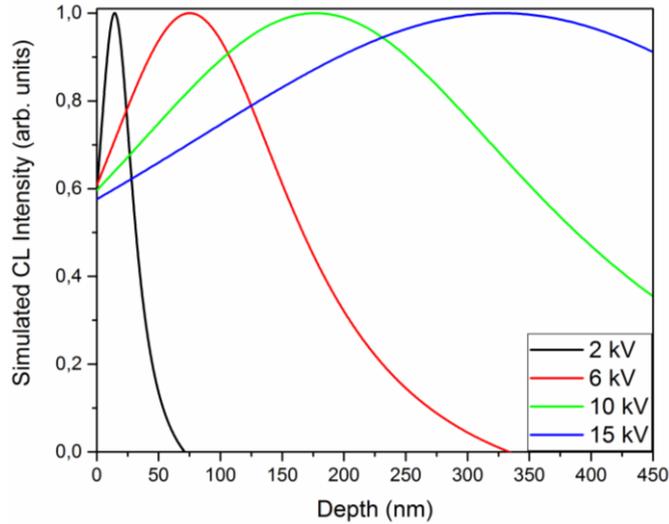

**Fig. 4.** The Monte Carlo simulations of the CL signal at different acceleration voltages. The results were normalized to 1 to make them easier to compare.

Fig. 5 compares the series of CL spectra for different AV (corresponding to different depths of the CL signal emission) and the distribution of Cd atoms measured by the SIMS are compared. The CL spectra are normalized to the maximum intensity to show the shift of the emission peak with changes in the Cd distribution in the SLs. At low temperatures, the near-band-gap blue emission dominates. For the $\{CdO_{12nm}/ZnO_{1.5nm}\}_{25}$ (A1) SL, the SIMS profile (Fig. 4) shows an increase in the Cd concentration in the deeper part of the structure. In low temperature (5 K) CL spectra excited by 2 keV electron beam (~ 20 nm depth) the emission peak at 3.191 eV due to the near-band-gap recombination is detected. The position of the main peak reveals a 32 meV red shift with increasing AV from 2 kV to 15 kV, which confirms an increase in the Cd content with depth, as shown by SIMS. In the CL signal from a deeper part of the structure also a second emission peak at 2.973 eV appears. This peak can be related either to the local Cd atom concentration or to defects. Abbasi at el. [49] suggested that at Cd content above 3% replacing the $Zn^{2+}$ ions with larger $Cd^{2+}$ ions created stress, and increased the number of defects.

It can be seen (Fig. 5), that the SIMS plot profile and CL spectra for the $\{CdO_{9.3nm}/ZnO_{4.5nm}\}_{18}$ (A2) structure are similar to those for the $\{CdO_{12nm}/ZnO_{1.5nm}\}_{25}$ (A1) structure. The position of the main peak shifts to lower energy from 3.151 eV to 3.143 eV with depth. Also for high kinetic energies of electrons (10 kV and 15 kV), a second emission peak is observed at lower energy. Apparently, the position of this peak reflects a higher concentration of Cd close to the substrate/SL interface.

The $\{CdO_{5nm}/ZnO_{5nm}\}_{25}$ (A3) and $\{ZnO_{9nm}/CdO_{5nm}\}_{18}$ (B2) SLs are characterized by a relatively uniform Cd distribution with depth, as revealed by the relatively flat profile on the SIMS plot.

Also, a nonuniform Cd distribution with depth is observed for the structure $\{ZnO_{8nm}/CdO_{1nm}\}_{25}$ (B1). The spectrum shows a strong redshift by about 60 meV.

The SIMS profile of the $\{ZnO_{5nm}/CdO_{5nm}\}_{25}$ (B3) SL reveals a relatively even distribution of cadmium. In the CL spectrum excited with 2 kV electrons two peaks at energies of 3.191 eV and 3.057 eV are observed. An increase of AV from 2 kV to 15 kV leads to a decrease in the intensity of the higher energy peak and its shift by 36 meV towards a lower energies and an increase in the intensity of the second peak and its blue shift by about 12 meV. Fig. 4 clearly shows that for a lower total Cd content in the SLs (thinner CdO layers), a more homogeneous distribution of Cd with depth is observed in the annealed samples.

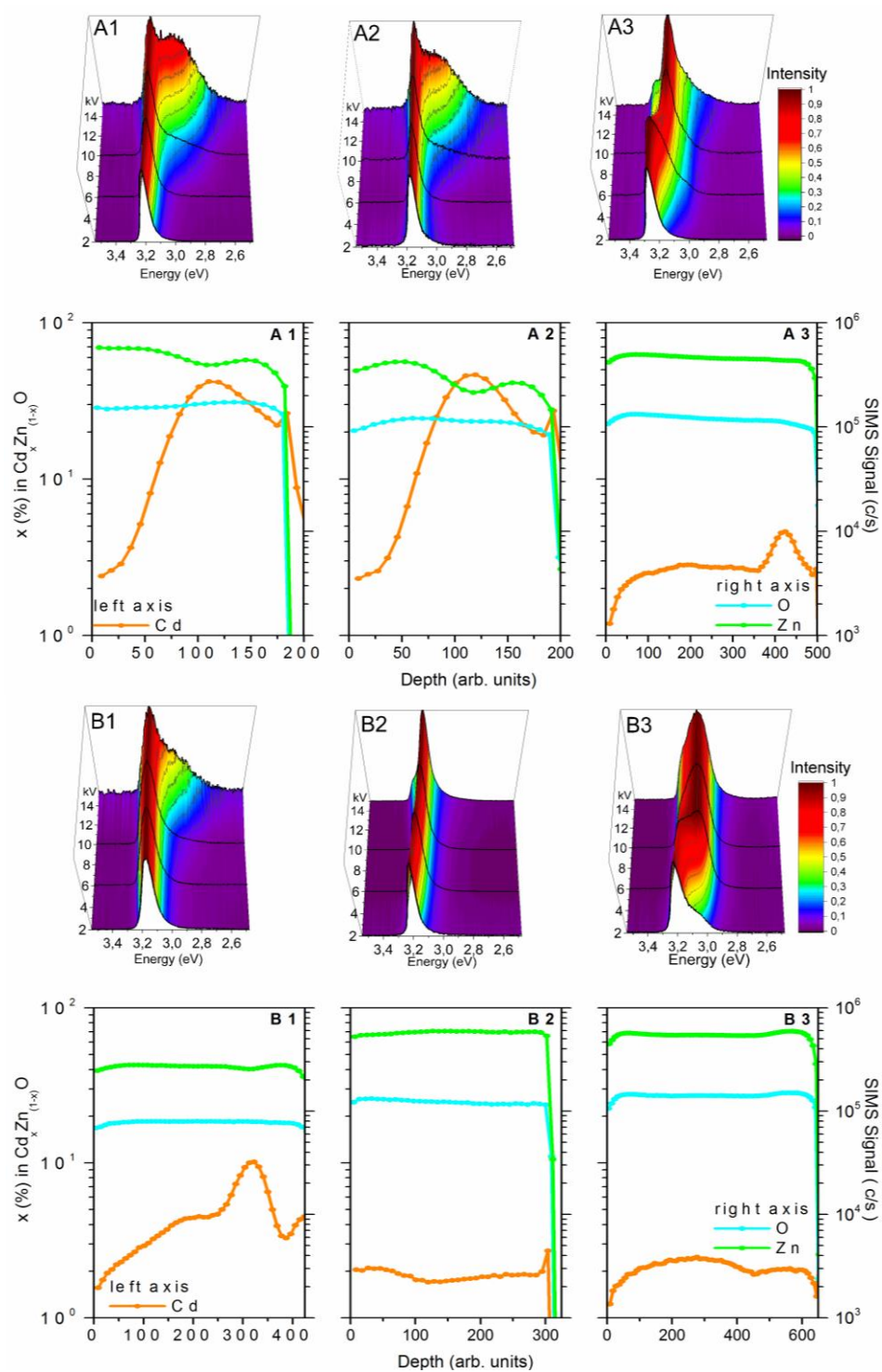

**Fig. 5.** The SIMS depth profiles of Cd, Zn and O elements in annealed {CdO/ZnO}$_m$ SLs compared to 3D CL spectra of {CdO/ZnO}$_m$ SLs at different electron beam kinetic energies.

To determine the value x in the annealed samples the cadmium implanted ZnO standard was measured. The Cd dose of $10^{16}$ cm$^{-2}$ at the energy of 250 keV allows to obtain the maximum concentration of $1.43 \times 10^{21}$ cm$^{-3}$, which correspond to x=3.45%. No matrix effect was observed in the SIMS data up to the Cd concentration measured in the implanted sample. Taking into account the results for PbSnSe [48] and the oxygen signal behavior in the SIMS depth profiles, it can be concluded that the matrix effect in the measurements of Cd content in ZnO is negligible up to x=30%.

The results of the CL studies in SLs are summarized in Fig. 6. As can be seen from Fig. 6, it is common for all structures that the Cd content near the surface is usually lower than inside the samples. Moreover, there is a one-to-one correspondence between the Cd depth profiles and the observed CL peak positions (Fig. 6 (a) and (b)). Cd segregation gives rise to different alloy compositions at different depths.

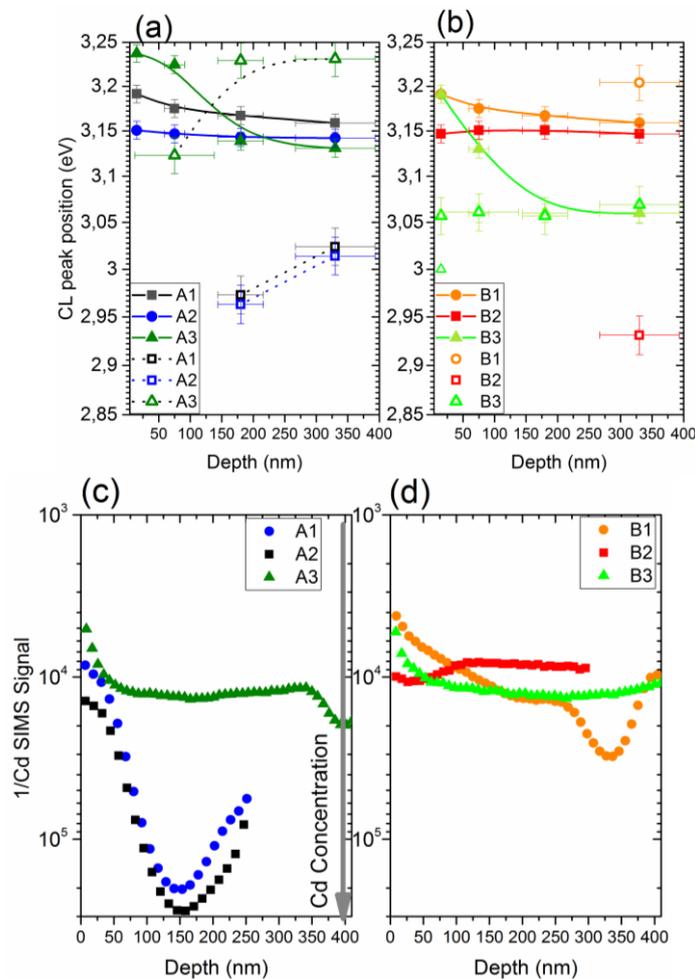

**Fig. 6.** (a) and (b) Peak positions vs depth calculated with CASINO software for annealed {CdO/ZnO}$_m$ structures; (c) and (d) are 1/Cd profiles.

**Conclusions:**

The formation of multilayer superlattice structures with different thicknesses of CdO and ZnO sublayers was confirmed by the SIMS analysis. As a result of high-temperature annealing, the well-defined initial crystal structure of the SLs is degraded. In particular, it was found that the thickness of the individual CdO and ZnO layers affects the final homogeneity of the Cd distribution in the annealed structures. As a consequence, an inhomogeneous distribution of Cd is often observed in the annealed structures. The SIMS depth profiles revealed segregation of Cd at the layer-substrate interface. The Cd profiles are also reflected in the CL spectra collected from different layer depths. Both experimental approaches (depth profiling using SIMS and depth-dependent CL) confirmed the presence of Cd- or Zn-rich regions in depth in some of the annealed samples.

**Acknowledgement**

This research was funded in whole or in part by National Science Center, Poland Grants No. 2019/35/B/ST8/01937, and 2021/41/B/ST5/00216. For the purpose of Open Access, the author has applied a CC-BY public copyright licence to any Author Accepted Manuscript (AAM) version arising from this submission.

**References**

[1] A. Stadler, Transparent Conducting Oxides—An Up-To-Date Overview, Materials (Basel). 5 (2012) 661–683. https://doi.org/10.3390/ma5040661.

[2] T. Minami, Transparent conducting oxide semiconductors for transparent electrodes, Semicond. Sci. Technol. 20 (2005) 34–44. https://doi.org/10.1088/0268-1242/20/4/004.

[3] T. Minami, New n-type transparent conducting oxides, MRS Bull. 25 (2000) 38–44. https://doi.org/10.1557/mrs2000.149.

[4] Y.H. Zhang, Z.X. Mei, H.L. Liang, X.L. Du, Review of flexible and transparent thin-film transistors based on zinc oxide and related materials, Chinese Phys. B. 26 (2017) 047307. https://doi.org/10.1088/1674-1056/26/4/047307.

[5] B.Y. Oh, M.C. Jeong, T.H. Moon, W. Lee, J.M. Myoung, J.Y. Hwang, D.S. Seo, Transparent conductive Al-doped ZnO films for liquid crystal displays, J. Appl. Phys. 99 (2006). https://doi.org/10.1063/1.2206417.

[6] D.C. Look, K.D. Leedy, ZnO plasmonics for telecommunications, Appl. Phys. Lett. 102 (2013) 182107. https://doi.org/10.1063/1.4804984.

[7] J. Klingshirn, C. F., Waag, A., Hoffmann, A., & Geurts, Zinc Oxide, Springer, 2010.

[8] E. Placzek-Popko, K.M. Paradowska, M.A. Pietrzyk, Z. Gumienny, P. Biegański, A. Kozanecki, Deep traps and photo-electric properties of p-Si/MgO/n-$Zn_{1-x}Mg_xO$ heterojunction, J. Appl. Phys. 118 (2015) 074501. https://doi.org/10.1063/1.4928728.

[9] R. Khokhra, B. Bharti, H.N. Lee, R. Kumar, Visible and UV photo-detection in ZnO nanostructured thin films via simple tuning of solution method, Sci. Rep. 7 (2017) 15032. https://doi.org/10.1038/s41598-017-15125-x.

[10] S.K. Hau, H.L. Yip, N.S. Baek, J. Zou, K. O'Malley, A.K.Y. Jen, Air-stable inverted flexible polymer solar cells using zinc oxide nanoparticles as an electron selective layer, Appl. Phys. Lett. 92 (2008) 253301. https://doi.org/10.1063/1.2945281.

[11] A. Kolodziejczak-Radzimska, T. Jesionowski, Zinc oxide-from synthesis to application: A review, Materials (Basel). 7 (2014) 2833–2881. https://doi.org/10.3390/ma7042833.

[12] Y. Choi, J. Kang, D. Hwang, S. Park, Recent Advances in ZnO-Based Light-Emitting Diodes, IEEE Trans. Electron Devices. 57 (2010) 26–41. https://doi.org/10.1109/TED.2009.2033769.

[13] E. Przezdziecka, A. Wierzbicka, P. Dłużewski, I. Sankowska, P. Sybilski, K. Morawiec, M.A. Pietrzyk, A. Kozanecki, Short-Period CdO/MgO Superlattices as Cubic CdMgO Quasi-Alloys, Cryst. Growth Des. 20 (2020) 5466–5472. https://doi.org/10.1021/acs.cgd.0c00678.

[14] J.F. Kong, W.Z. Shen, Y.W. Zhang, C. Yang, X.M. Li, Resonant Raman scattering probe of alloying effect in ZnMgO thin films, Appl. Phys. Lett. 92 (2008) 191910. https://doi.org/10.1063/1.2930676.

[15] L. Su, Y. Zhu, Q. Zhang, M. Chen, T. Wu, X. Gui, B. Pan, R. Xiang, Z. Tang, Structure and optical properties of ternary alloy BeZnO and quaternary alloy BeMgZnO films growth by molecular beam epitaxy, Appl. Surf. Sci. 274 (2013) 341–344. https://doi.org/10.1016/j.apsusc.2013.03.058.

[16] Y.R. Ryu, T.S. Lee, J.A. Lubguban, A.B. Corman, H.W. White, J.H. Leem, M.S. Han, Y.S. Park, C.J. Youn, W.J. Kim, Wide-band gap oxide alloy: BeZnO, Appl. Phys. Lett. 88 (2006) 052103. https://doi.org/10.1063/1.2168040.

[17] K.P. Misra, R.K. Shukla, A. Srivastavaa, A. Srivastava, Blueshift in optical band gap in nanocrystalline $Zn_{1-x}Ca_xO$ films deposited by sol-gel method, Appl. Phys. Lett. 95 (2009) 031901. https://doi.org/10.1063/1.3184789.

[18]  A. Fouzri, N.A. Althumairi, V. Sallet, A. Lusson, Characterization of sol gel $Zn_{1-x}Ca_xO$ thin layers deposited on p-Si substrate by spin-coating method, Opt. Mater. (Amst). 110 (2020) 110519. https://doi.org/10.1016/j.optmat.2020.110519.

[19]  C. Yang, X.M. Li, Y.F. Gu, W.D. Yu, X.D. Gao, Y.W. Zhang, ZnO based oxide system with continuous bandgap modulation from 3.7 to 4.9 eV, Appl. Phys. Lett. 93 (2008) 112114. https://doi.org/10.1063/1.2987420.

[20]  J. Zúñiga-Pérez, ZnCdO: Status after 20 years of research, Mater. Sci. Semicond. Process. 69 (2017) 36–43. https://doi.org/10.1016/j.mssp.2016.12.002.

[21]  S. Mondal, S. Paul, M.J. Alam, S. Sushama, S. Chakrabarti, Effects of carrier confinement in MgZnO/CdZnO thin-film transistors: Towards next generation display technologies, Superlattices Microstruct. 134 (2019) 106220. https://doi.org/10.1016/j.spmi.2019.106220.

[22]  R. Vettumperumal, S. Kalyanaraman, R. Thangavel, Photoconductive UV detectors based heterostructures of Cd and Mg doped ZnO sol gel thin films, Mater. Chem. Phys. 145 (2014) 237–242. https://doi.org/10.1016/j.matchemphys.2014.02.008.

[23]  V. Kaushik, S. Rajput, M. Kumar, Broadband optical modulation in a zinc-oxide-based heterojunction via optical lifting, Opt. Lett. 45 (2020) 363. https://doi.org/10.1364/ol.379257.

[24]  Y.Z. Zhu, G.D. Chen, H. Ye, A. Walsh, C.Y. Moon, S.H. Wei, Electronic structure and phase stability of MgO, ZnO, CdO, and related ternary alloys, Phys. Rev. B - Condens. Matter Mater. Phys. 77 (2008) 245209. https://doi.org/10.1103/PhysRevB.77.245209.

[25]  H. Rozale, B. Bouhafs, P. Ruterana, First-principles calculations of the optical band-gaps of $Zn_xCd_{1-x}O$ alloys, Superlattices Microstruct. 42 (2007) 165–171. https://doi.org/10.1016/j.spmi.2007.04.007.

[26]  X. Ma, P. Chen, R. Zhang, D. Yang, Optical properties of sputtered hexagonal CdZnO films with band gap energies from 1.8 to 3.3 eV, J. Alloys Compd. 509 (2011) 6599–6602. https://doi.org/10.1016/j.jallcom.2011.03.101.

[27]  D.M. Detert, S.H.M. Lim, K. Tom, A. V. Luce, A. Anders, O.D. Dubon, K.M. Yu, W. Walukiewicz, Crystal structure and properties of $Cd_xZn_{1-x}O$ alloys across the full composition range, Appl. Phys. Lett. 102 (2013) 232103. https://doi.org/10.1063/1.4809950.

[28]  H. Köhler, Optical properties and energy-band structure of CdO, Solid State Commun. 11 (1972) 1687–1690. https://doi.org/10.1016/0038-1098(72)90772-7.

[29]  M. Lange, C.P. Dietrich, G. Benndorf, M. Lorenz, J. Zúñiga-Pérez, M. Grundmann, Thermal stability of ZnO/ZnCdO/ZnO double heterostructures grown by pulsed laser deposition, J. Cryst. Growth. 328 (2011) 13–17. https://doi.org/10.1016/j.jcrysgro.2011.06.030.

[30]  G. Yao, Y. Tang, Y. Fu, Z. Jiang, X. An, Y. Chen, Y. Liu, Fabrication of high-quality ZnCdO epilayers and ZnO/ZnCdO heterojunction on sapphire substrates by pulsed laser deposition, Appl. Surf. Sci. 326 (2015) 271–275. https://doi.org/10.1016/j.apsusc.2014.11.045.

[31]  O. Aguilar, S. de Castro, M.P.F. Godoy, M. Rebello Sousa Dias, Optoelectronic characterization of $Zn_{1-x}Cd_xO$ thin films as an alternative to photonic crystals in organic solar cells, Opt. Mater. Express. 9 (2019) 3638-3648. https://doi.org/10.1364/ome.9.003638.

[32]  L.P Purohit, Comparative Study of Synthesis of CdO-ZnO Nanocomposite Thin Films by Different Methods: A Review, Nanosci. Technol. Open Access. 3 (2016) 1–5. https://doi.org/10.15226/2374-8141/3/2/00140.

[33]  S. Vijayalakshmi, S. Venkataraj, R. Jayavel, Characterization of cadmium doped zinc

oxide (Cd : ZnO) thin films prepared by spray pyrolysis method, J. Phys. D. Appl. Phys. 41 (2008) 245403. https://doi.org/10.1088/0022-3727/41/24/245403.

[34] Y.S. Choi, C.G. Lee, S.M. Cho, Transparent conducting $Zn_xCd_{1-x}O$ thin films prepared by the sol-gel process, Thin Solid Films. 289 (1996) 153–158. https://doi.org/10.1016/S0040-6090(96)08923-7.

[35] S. Gowrishankar, L. Balakrishnan, N. Gopalakrishnan, Band gap engineering in $Zn_{(1-x)}Cd_xO$ and $Zn_{(1-x)}Mg_xO$ thin films by RF sputtering, Ceram. Int. 40 (2014) 2135–2142. https://doi.org/10.1016/j.ceramint.2013.07.130.

[36] G. V. Lashkarev, I.I. Shtepliuk, A.I. Ievtushenko, O.Y. Khyzhun, V. V. Kartuzov, L.I. Ovsiannikova, V.A. Karpyna, D. V. Myroniuk, V. V. Khomyak, V.N. Tkach, I.I. Timofeeva, V.I. Popovich, N. V. Dranchuk, V.D. Khranovskyy, P. V. Demydiuk, Properties of solid solutions, doped film, and nanocomposite structures based on zinc oxide, Low Temp. Phys. 41 (2015) 129–140. https://doi.org/10.1063/1.4908204.

[37] H.C. Jang, K. Saito, Q. Guo, K.M. Yu, W. Walukiewicz, T. Tanaka, Realization of rocksalt $Zn_{1-x}Cd_xO$ thin films with an optical band gap above 3.0 eV by molecular beam epitaxy, CrystEngComm. 22 (2020) 2781–2787. https://doi.org/10.1039/c9ce02018g.

[38] S. Sadofev, S. Blumstengel, J. Cui, J. Puls, S. Rogaschewski, P. Schäfer, F. Henneberger, Visible band-gap ZnCdO heterostructures grown by molecular beam epitaxy, Appl. Phys. Lett. 89 (2006) 201907. https://doi.org/10.1063/1.2388250.

[39] V. Venkatachalapathy, A. Galeckas, M. Trunk, T. Zhang, A. Azarov, A.Y. Kuznetsov, Understanding phase separation in ZnCdO by a combination of structural and optical analysis, Phys. Rev. B - Condens. Matter Mater. Phys. 83 (2011) 125315. https://doi.org/10.1103/PhysRevB.83.125315.

[40] K. Sakurai, T. Kubo, D. Kajita, T. Tanabe, H. Takasu, S. Fujita, S. Fujita, Blue photoluminescence from ZnCdO films grown by molecular beam epitaxy, Jpn. J. Appl. Phys. 39 (2000) L 1146 – L 1148. https://doi.org/10.1143/jjap.39.l1146.

[41] A. V. Thompson, C. Boutwell, J.W. Mares, W. V. Schoenfeld, A. Osinsky, B. Hertog, J.Q. Xie, S.J. Pearton, D.P. Norton, Thermal stability of CdZnO/ZnO multi-quantum-wells, Appl. Phys. Lett. 91 (2007) 201921. https://doi.org/10.1063/1.2812544.

[42] L. Li, Z. Yang, Z. Zuo, J.H. Lim, J.L. Liu, Thermal stability of CdZnO thin films grown by molecular-beam epitaxy, Appl. Surf. Sci. 256 (2010) 4734–4737. https://doi.org/10.1016/j.apsusc.2010.02.083.

[43] A.Y. Azarov, T.C. Zhang, B.G. Svensson, A.Y. Kuznetsov, Cd diffusion and thermal stability of CdZnO/ZnO heterostructures, Appl. Phys. Lett. 99 (2011) 111903. https://doi.org/10.1063/1.3639129.

[44] M.A. Pietrzyk, A. Wierzbicka, E. Zielony, A. Pieniazek, R. Szymon, E. Placzek-Popko, Fundamental studies of ZnO nanowires with ZnCdO/ZnO multiple quantum wells grown for tunable light emitters, Sensors Actuators, A Phys. 315 (2020) 112305. https://doi.org/10.1016/j.sna.2020.112305.

[45] I. Suemune, A.B.M.A. Ashrafi, M. Ebihara, M. Kurimoto, H. Kumano, T.Y. Seong, B.J. Kim, Y.W. Ok, Epitaxial ZnO growth and p-type doping with MOMBE, Phys. Status Solidi Basic Res. 241 (2004) 640–647. https://doi.org/10.1002/pssb.200304290.

[46] E. Przezdziecka; A. Wierzbicka; A. Lysak; P. Dłużewski; A. Adhikari; P. Sybilski; K. Morawiec; A. Kozanecki, Nanoscale morphology of short-period {CdOZnO} superlattices grown by MBE, Cryst. Growth Des. (2022).

[47] D. Drouin, A.R. Couture, D. Joly, X. Tastet, V. Aimez, R. Gauvin, CASINO V2.42 - A fast and easy-to-use modeling tool for scanning electron microscopy and microanalysis users, Scanning. 29 (2007) 92–101. https://doi.org/10.1002/sca.20000.

[48] R. Jakieła, M. Galicka, P. Dziawa, G. Springholz, A. Barcz, SIMS accurate determination of matrix composition of topological crystalline insulator material $Pb_{1-x}Sn_xSe$, Surf. Interface Anal. 52 (2020) 71–75. https://doi.org/10.1002/sia.6705.

[49] F. Abbasi, F. Zahedi, Performance improvement of UV photodetectors using Cd-doped ZnO nanostructures, Res. Sq. https://doi.org/10.21203/rs.3.rs-328527/v1.